\documentclass[reqno,12pt]{amsart}
\usepackage{fullpage}
\usepackage{amsfonts}
\usepackage{amssymb}

\def\ov{\overline}

\def\pe{\perp}
\def\pa{\parallel}

\def\mb{\mathbf}
\def\mr{\mathrm}
\def\mk{\mathfrak}
\def\bs{\boldsymbol}
\def\tb{\mathbf}

\def\Rnum{\mathbb{R}}
\def\Cnum{\mathbb{C}}

\def\i{\mr{i}}

\def\Re{{\rm Re}}
\def\Im{{\rm Im}}
\def\id{{\rm id}}

\def\inv{{}^{-1}}
\def\t{{\rm t}}

\def\<{\langle}
\def\>{\rangle}

\def\ha{\tfrac{1}{2}}

\def\ad{{\rm ad}}
\def\Ad{{\rm Ad}}

\def\tr{{\rm tr}}

\def\msp{\mk{m}}
\def\hsp{\mk{h}}
\def\gsp{\mk{g}}

\def\ee{\mr{e}}

\def\conx{\omega}

\def\cvec{\mb{c}}
\def\avec{\mb{a}}
\def\bvec{\mb{b}}

\def\uvec{\mb{u}}
\def\hvec{\mb{h}}
\def\wvec{\mb{w}}

\def\H{\mb{H}}
\def\W{\mb{W}}

\def\A{\mb{A}}
\def\C{\mb{C}}

\def\u{\mr{u}}
\def\w{\mr{w}}
\def\h{\mr{h}}

\def\a{\mr{a}}

\def\c{\mr{c}}

\def\Rop{{\mathcal R}}
\def\Jop{{\mathcal J}}
\def\Hop{{\mathcal H}}

\def\Kop{{\mathcal K}}

\def\map{\gamma}

\def\secref#1{Sec.~\ref{#1}}

\def\c{\mr{c}}

\def\spn{\mk{sp}(n)}
\def\un{\mk{u}(n)}
\def\bel{\begin{lemma}}
\def\eel{\end{lemma}}
\def\ber{\begin{remark}}
\def\eer{\end{remark}}
\def\bep{\begin{proposition}}
\def\eep{\end{proposition}}
\def\bpm{\begin{pmatrix}}
\def\epm{\end{pmatrix}}
\def\beq{\begin{equation}}
\def\eeq{\end{equation}}

\newtheorem{theorem}{Theorem}{\bf}{\it}

\newtheorem{lemma}[theorem]{Lemma}{\bf}{\it}

{\bf}{\it}

\newtheorem{proposition}[theorem]{Proposition}{\bf}{\it}

\begin{document}


\title{Integrable Systems with Unitary Invariance from Non-stretching Geometric Curve Flows in the Hermitian Symmetric Space $Sp(n)/U(n)$}

\author{STEPHEN C. ANCO}

\address{Department of Mathematics and Statistics, Brock University\\
St. Catharines, Canada \\
sanco@brocku.ca}

\author{ESMAEEL ASADI}

\address{Department of Mathematics, Institute for Advanced Studies in Basic Sciences\\
Zanjan 45137-66731,
Iran\\
easadi@iasbs.ac.ir}

\author{ASIEH DOGONCHI}

\address{Department of Mathematics, Institute for Advanced Studies in Basic Sciences \\
Zanjan 45137-66731,
Iran\\
a.dogonchi@iasbs.ac.ir}

\maketitle

\begin{abstract}
A moving parallel frame method is applied to geometric non-stretching curve flows 
in the Hermitian symmetric space $Sp(n)/U(n)$ 
to derive new integrable systems with unitary invariance. 
These systems consist of a bi-Hamiltonian modified Korteweg-de Vries equation 
and a Hamiltonian sine-Gordon (SG) equation, 
involving a scalar variable coupled to a complex vector variable. 
The Hermitian structure of the symmetric space $Sp(n)/U(n)$ is used 
in a natural way from the beginning in formulating 
a complex matrix representation of the tangent space $\mk{sp}(n)/\mk{u}(n)$ 
and its bracket relations within the symmetric Lie algebra $(\mk{u}(n),\mk{sp}(n))$.
\end{abstract}


\section{Introduction}

In the theory of integrable systems, 
the modified Korteweg-de Vries (mKdV) equation 
and the sine-Gordon (SG) equation 
are two integrable equations of basic importance. 
They have an elegant geometric origin that arises from 
the differential invariant of a curve that undergoes 
certain geometric non-stretching flows 
(which locally preserve the arclength of the curve) 
in the Euclidean plane and the sphere, respectively. 
For generalizing this geometric picture to higher dimensional spaces,
it becomes natural to transform from a Frenet frame along the curve 
which determines the differential invariants of the curve 
to a parallel frame \cite{Bishop} 
which determines differential covariants of the curve.
This approach has been used 
in numerous geometric spaces
(see Refs.~\cite{Hasimoto,DoliwaSantini,LangerPerline,LangerPerline00,SandersWang,Sanders.Wang.Beffa,AncoJPA,AncoSIGMA,AncoIMA,AsadiSanders,AsadiPhDthesis,AncoAsadi2009,AncoAsadi2012}). 

There is a general construction of a parallel frame \cite{AncoJGP}
for non-stretching curve flows in Riemannian symmetric spaces $M=G/H$. 
The Cartan structure equations of this frame have the property 
that they explicitly yield a pair of Hamiltonian and symplectic operators 
in all symmetric spaces. 
These operators can be used to derive a hierarchy of group-invariant 
bi-Hamiltonian mKdV equations and SG equations
whose variables are given by the components of 
the Cartan connection matrix along a curve undergoing 
particular geometric non-stretching flows \cite{AncoJGP}.

In this paper, we apply this general method \cite{AncoJGP} 
to the Hermitian symmetric space $Sp(n)/U(n)$ 
and get $U(n-1)$-invariant integrable scalar-vector mKdV and SG equations. 
These equations and their bi-Hamiltonian integrability structure 
naturally involve both the Hermitian inner product 
and the complex structure $J$ on the symmetric Lie algebra 
$(\mk{u}(n),\mk{sp}(n)/\mk{u}(n))$. 

Using this Hermitian structure, 
we show how to introduce a natural complex matrix representation 
for the symmetric Lie algebra $(\mk{u}(n),\mk{sp}(n)/\mk{u}(n))$ 
as well as the Lie bracket relations between $\mk{u}(n)$ and $\mk{sp}(n)/\mk{u}(n))$.  
In this way, the flow equation can be naturally written in terms of
a scalar variable and a complex vector variable that transform properly
under the unitary gauge group of the underlying parallel frame. 
If instead a standard real matrix representation were to be used, 
there would be a problem of how to translate the flow equation with real variables
into a unitarily invariant form. 
Our approach fully resolves this issue. 
(See also the examples in Refs.~\cite{AncoIMA,AncoAsadi2009,AncoAsadi2012}.)

More generally, depending on whether the symmetric space is 
real, Hermitian, or quaternionic, 
one may use such extra geometric/algebraic structures 
from the beginning on the tangent space of $M$
as well as on the associated symmetric Lie algebra $(\hsp,\gsp/\hsp)$
and its Lie bracket relations. 

The outline of the paper is as follows. 
In section 2, we review the definition of Riemannian symmetric spaces 
and also Hermitian symmetric spaces 
using symmetric Lie algebras. 
For the Hermitian symmetric space $Sp(n)/U(n)$, 
we decompose the subalgebra $\mk{u}(n)$ 
and the quotient space $\mk{sp}(n)/\mk{u}(n)$ 
into parallel and perp subspaces relative to a regular element in the Cartan 
subspace in $\mk{sp}(n)/\mk{u}(n)$. 
We give a complex matrix representation of these subspaces 
and also the bracket relations between them,
which will be essential later for setting up a parallel frame 
for curves in $Sp(n)/U(n)$. 

In section 3, 
we first review the standard frame field and connection formulation of 
the Riemannian structure common to all symmetric spaces,
and then we review the definition of a moving parallel frame 
for non-stretching curve flows in symmetric spaces. 
In section 4, 
we consider non-stretching curve flows in the Hermitian space $M=Sp(n)/U(n)$. 
By utilizing a moving parallel frame and then taking the pull back of the torsion and curvature $2$-forms from $M$ to the surface swept out by a curve flow, 
we get Cartan structure equations which encode a pair of Hamiltonian and symplectic operators. 
These results will be just stated without proof. 
See Ref.~\cite{AncoJGP} for details and an explicit proof for curve flows in general symmetric spaces. 
Then we derive a hierarchy of $U(n-1)$-invariant bi-Hamiltonian equations,
starting with a $U(n-1)$-invariant mKdV equation. 
All of these equations are integrable systems for 
an imaginary scalar variable coupled to a complex vector variable. 
We next derive a $U(n-1)$-invariant SG equation for the same variables 
by using the kernel of the symplectic operator and doing an algebraic reduction of the parallel part of the flow. 
There is a general method behind this reduction which will be explored in a subsequent paper\cite{Anco.Asadi.Reduction}.

\section{Algebraic structure of $Sp(n)/U(n)$}

A general Riemannian symmetric space $M=G/H$ 
is defined by \cite{Helgason} 
a simple Lie group $G$ and an involutive compact Lie subgroup $H$ in $G$. 
Any linear frame on $M$ provides a soldering identification 
\cite{KobayashiNomizu} 
between the tangent space $T_xM$ at points $x$
and the vector space $\msp=\gsp/\hsp$.
There is an orthogonal decomposition $\gsp=\hsp\oplus\msp$ 
with respect to the Cartan-Killing form with the Lie bracket relations
\beq\label{ber.rel}
[\hsp,\hsp]\subset \hsp,
\quad
[\hsp,\msp] \subset \msp,
\quad [\msp,\msp]\subset \hsp.
\eeq

A Hermitian symmetric space is a Riemannian symmetric space that possesses
a covariantly-constant $(1,1)$ tensor $J$ which acts on each tangent space 
as a complex structure, $J^2=-\id$. 

Cartan subspaces of $\msp$ are defined as a maximal abelian subspace
$\mk{a}\subseteq\msp$ 
with the property that it is the centralizer of its elements,
$\mk{a}=\msp\cap\mk{c}(\mk{a})$.
It is well-known \cite{Helgason}
that any two Cartan subspaces are isomorphic to one another
under some linear transformation in $\Ad(H)$
and that the action of the linear transformation group $\Ad(H)$
on any Cartan subspace $\mk{a}$ generates $\msp$.
The dimension of $\mk{a}$ as a vector space is equal to the rank of $M$.

To define a moving parallel frame on non-stretching curve flows,
which will be presented in the next section, 
we need to choose an element $\ee$ in the Cartan subspace $\mk{a}$. 
This element will be identified with the frame components of 
the tangent vector of a curve. 
For any choice of $\ee$, 
the corresponding linear operator $\ad(\ee)$
induces a direct sum decomposition of the vector spaces
$\msp=\mk{sp}(n)/\mk{u}(n)$ and $\hsp= \mk{u}(n)$
into centralizer spaces $\msp_{\pa}$ and $\hsp_{\pa}$
and their orthogonal complements (perp spaces)
$\msp_{\pe}$ and $\hsp_{\pe}$
with respect to the Cartan-Killing form. The Lie bracket relations on
$\msp_{\pa}$, $\msp_{\pe}$, $\hsp_{\pa}$, $\hsp_{\pe}$
coming from the structure of $\gsp$
as a symmetric Lie algebra \eqref{ber.rel} are given by 
\begin{align}
&
[\msp_\pa,\msp_\pa] \subseteq \hsp_\pa,
\quad
[\msp_\pa,\hsp_\pa] \subseteq \msp_\pa,
\quad
[\hsp_\pa,\hsp_\pa] \subseteq \hsp_\pa,
\label{inclusion.one}\\
&
[\hsp_\pa,\msp_\pe] \subseteq \msp_\pe,
\quad
[\hsp_\pa,\hsp_\pe] \subseteq \hsp_\pe,
\label{inclusion.two}\\
&
[\msp_\pa,\msp_\pe] \subseteq \hsp_\pe,
\quad
[\msp_\pa,\hsp_\pe] \subseteq \msp_\pe,
\label{inclusion.three}
\end{align}
while the remaining Lie brackets
\beq
[\msp_\pe,\msp_\pe],
\quad
[\hsp_\pe,\hsp_\pe],
\quad
[\msp_\pe,\hsp_\pe]
\label{inclusion.gen}
\eeq
obey the general relations \eqref{ber.rel}.

Through the Lie bracket relations \eqref{inclusion.one}--\eqref{inclusion.three},
the operator $\ad(\ee)$ maps $\hsp_{\pe}$ into $\msp_{\pe}$,
and vice versa.
Hence $\ad(\ee)^2$ is well-defined as a linear mapping of
each subspace $\hsp_{\pe}$ and $\msp_{\pe}$ into itself.

\subsection{Symmetric Lie algebra $(\mk{u}(n),\mk{sp}(n)/\mk{u}(n))$}
\label{g=un}

Hermitian symmetric spaces have been classified by Cartan and others. 
See Ref.~\cite{Helgason} for full details.
For the specific Hermitian symmetric space $Sp(n)/U(n)$, 
we will provide all the algebraic details needed for later use.

The symplectic Lie algebra $\mk{sp}(n)$ consists of all matrices 
in $\mk{gl}(2n,\Cnum)$ satisfying the condition 
\begin{equation}
g\theta+\theta g^t=0,
\quad 
g^t=-\ov g,
\quad 
\theta=
\bpm
  0&I\\
 -I&0
\epm.
\end{equation}
An involution $\sigma$ on $\mk{gl}(2n,\Cnum)$ 
which leaves $\mk{sp}(n)$ invariant is defined by $\sigma(g)=\ov g$.
The set of matrices in $\mk{sp}(n)$ that is invariant under $\sigma$ is isomorphic to the unitary Lie algebra $\mk{u}(n)$. 
Thus the Lie algebra $\mk{sp}(n)$ as a symmetric Lie algebra  
orthogonally decomposes as $\mk{sp}(n)=\hsp\oplus\msp$ 
into the eigenspaces of $\sigma$:
\begin{align}  
\hsp:=&\mk{u}(n)\subset \gsp,
\quad 
\sigma(\hsp)=\hsp,  
\\
\msp:=&\mk{sp}(n)/\mk{u}(n)\subset\gsp,
\quad 
\sigma(\msp)=-\msp.
\end{align}
  
The matrix representation of $\gsp=\mk{sp}(n)$ is given as 
\beq
\bpm
    A & B\\
    -\ov B & \ov A
\epm
\in \mk{sp}(n),
\quad 
A^t=-\ov{A},
\quad 
B^t=B
\eeq
in which  $A,B\in \mk{gl}(n,\Cnum)$. 
The matrix representation of the vector spaces 
$\msp=\mk{sp}(n)/\mk{u}(n)$ and $\hsp=\mk{u}(n)$ is given as
\begin{align}
   (A,B):=\i\begin{pmatrix}
     A& B\\
     B& - A
     \end{pmatrix}\in \msp,\quad A,B \in \mk{gl}(n,\Rnum),\quad A^t=A,\quad B^t=B,
\\
     (C,D):=\begin{pmatrix}
      C& -D\\
      D& C
      \end{pmatrix}\in  \hsp,\quad C,D\in \mk{gl}(n,\Rnum),\quad C^t=-C,\quad D^t=D.
\end{align}
 
The complex structure $J$ of the tangent space $T_oM=\msp$ 
is represented by $J=\ha \i I\in\mk{u}(n)$, 
so then the action of $\ad(J)$ on $\msp$ is simply given by multiplication 
by $\i$. 
Therefore one can identify $(A,B)\in\msp$ as a complex symmetric matrix $A+\i B\in\mk{s}(n,\Cnum)$. 
The elements $(C,D)$ in $\hsp$ can be naturally written as a complex matrix $C+\i D\in\mk{u}(n)$. 

With this complex representation, the bracket relation \eqref{ber.rel} 
is not just a matrix commutator due to the role of complex structure $J$. 

\bel\label{alg-vec.sp-un}
\begin{enumerate}
\item
The Hermitian matrix representation of the vector space
$\msp=\mk{sp}(n)/\mk{u}(n)$ and Lie subalgebra  $\hsp=\mk{u}(n)$  
is given as 
\begin{align}
&(A_1)\in \msp,\quad A_1\in\mk{gl}(n,\Cnum),\quad A_1^\t-A_1=0,\\
& (C_1)\in  \hsp,\quad C_1\in \mk{gl}(n,\Cnum),\quad \ov{C}_1^\t+C_1=0.
\end{align}
\item 
The Lie bracket relations \eqref{ber.rel} 
have the following matrix representation in which 
$(A_1),(A_2)\in \msp$ and $(C_1),(C_2)\in \hsp$:
\begin{align}
  [A_1,A_2]&=A_2\ov A_1-A_1\ov A_2\in \hsp, \\
   [A_1,C_1] &=A_1\ov C_1-C_1A_1 \in \msp, \\
    [C_1,C_2] &=C_1C_2-C_2C_1 \in \hsp.
\end{align}
\item
The restriction of Cartan-Killing form on $\mk{sp}(n)$ 
to $\msp=\mk{sp}(n)/\mk{u}(n)$ is a negative-definite inner product
\beq
\langle A_1,A_2\rangle=8(n+1)\Re(\tr(A_1\ov A_2)). 
\eeq
\item 
The vector space $\msp=\mk{sp}(n)/\mk{u}(n)$ is of dimension $ n^2+n $ and of rank $n$.  
The $n$-dimensional vector subspace 
$\mk{a} \subset \msp=\spn/\un$ generated by real, diagonal matrices 
$(E) \in\msp$ is a Cartan subspace.
\end{enumerate}
\eel

An element $\ee$ of the Cartan subspace $\mk{a}$ 
is called a {\it regular element} \cite{Helgason,Cartan.sub}
if its centralizer subspace $\mk{c}(\ee)$ in $\gsp=\mk{sp}(n)$ 
is of maximal dimension. 
Any real, diagonal matrix $E_{ii}\in\mk{gl}(n,\Rnum)$ 
whose only non-zero component is a $1$ in its $i$th row and $i$th column 
(with $1\leq i\leq n$) is regular element $(E_{ii})\in\mk{a}$. 
We will choose 
\beq\label{ee}
\ee:=\frac{1}{\sqrt{\chi}}(E_{11})\in \mk{a} 
\eeq
where $\chi\in\Rnum$ is a normalization constant. 
We choose this constant so that $\ee$ has unit norm,
$$ -1=\langle \ee,\ee \rangle=-8(n+1)/\chi $$
which determines
\beq
\chi=8(n+1).
\eeq

In the following lemma, we give an explicit matrix representation of 
the perp and parallel subspaces of $\msp$ and $\hsp$ determined by $\ee$.
These subspaces are defined by the properties
$$
\ad(\ee)\msp_\pa=0,
\quad
\langle \msp_\pe,\msp_\pa \rangle =0
$$
and
$$
\ad(\ee)\hsp_\pa=0,
\quad
\langle \hsp_\pe,\hsp_\pa \rangle =0 .
$$

\bel
\begin{enumerate}
\item 
The matrix representation of
$\msp_\pa$
and 
$\msp_\pe$
in $\msp=\spn/\un $
is given as 
\beq
( \a_{\pa},\A_\pa)=
\bpm
\a_\pa & 0\\
0 & \A_\pa
\epm
\in \msp_\pa,
\quad 
(\a_\pe,\avec_\pe):=
\bpm
\a_\pe & \avec_\pe\\
\avec_\pe^\t & 0
\epm 
\in \msp_\pe\label{pres.m-pa.un}
\eeq
in which 
$\a_\pa\in\Rnum$, $\a_\pe\in \i\Rnum$, $\avec_\pe\in \Cnum^{n-1}$, 
and $\A_\pa\in\mk{s}(n-1,\Cnum)$.
\item
The matrix representation of
$\hsp_\pa$
and
$\hsp_\pe$
is given as 
\begin{equation}\label{pres.h-pa.un}
(\C_\pa)=
\bpm
0 & 0\\
0 & \C_\pa
\epm
\in \hsp_\pa ,
\quad 
(\c_\pe,\cvec_\pe)=
\bpm
\c_\pe & \cvec_\pe\\
-\ov{\cvec}_\pe^\t & 0
\epm
\in\hsp_\pe
\end{equation}
in which
$\c_\pe \in \i\Rnum$, $\cvec_\pe\in \Cnum^{n-1}$, 
and $\C_\pa\in \mk{u}(n-1)$.
\item
The dimension of perp and parallel subspaces is given as 
\beq\begin{aligned}
&\dim(\msp_{\pa})=n^2-n+1,\\
&\dim(\hsp_{\pa})=n^2-2n+1,\\
&\dim(\msp_{\pe})=\dim(\hsp_{\pe})=2n-1.
\end{aligned}
\eeq
\item
The regular element \eqref{ee} in the Cartan subspace $\mk{a}$
is represented as
\beq\label{e}
\ee=\frac{1}{\sqrt{\chi}}(1,0)\in\msp_{\pa}.
\eeq
In particular the linear operator $\ad(\ee)$
gives an isomorphism of $\msp_\pe$ and $\hsp_\pe$:
\beq\label{sp.ad.e.h.pe} 
\begin{aligned} 
&  \ad(\ee)(\a_\pe,\avec_\pe)=
 \frac{1}{\sqrt{\chi}} (2\a_\pe,-\ov {\avec}_\pe)
\in \hsp_\pe, 
\\
& \ad(\ee)(\c_\pe,\cvec_\pe) =\frac{1}{\sqrt{\chi}}(-2\c_\pe,\ov{\cvec}_\pe)
\in \msp_\pe .
\end{aligned}
\eeq
\end{enumerate} 
\eel
   
To write out the explicit Lie bracket relations on
$\msp=\msp_{\pa}\oplus\msp_{\pe}$
and $\hsp=\hsp_{\pa}\oplus\hsp_{\pe}$,
we will use the following inner products and outer products.
For $\avec,\bvec\in\Cnum^{n-1}$,
we note
\beq\label{anti.conj.vec}
\begin{aligned}
&\ov{\avec}\bvec^\t + \ov{\bvec}\avec^\t
=2\Re\langle\avec,\bvec\rangle \in\Rnum,
\\
&\ov{\avec}\bvec^\t - \ov{\bvec}\avec^\t
=\i 2\Im\langle\avec,\bvec\rangle \in\i\Rnum,
\end{aligned}
\eeq
where 
\beq
\langle\avec,\bvec\rangle=\ov{\avec}\bvec^\t
\eeq
is the Hermitian inner product.
Also, we define
\beq\label{anti.conj.mat}
\begin{aligned}
&\avec^\t\bvec-\bvec^\t\avec:= \avec\wedge\bvec
\in\mk{so}(n-1,\Cnum),\\
&\avec^\t\bvec+\bvec^\t\avec:= \avec\odot\bvec
\in\mk{s}(n-1,\Cnum),
\end{aligned}
\eeq
and
\beq
\begin{aligned}
&\ov{\avec}^\t\bvec - \ov{\bvec}^\t\avec^\t:= \avec \ov{\wedge}\bvec 
\in\mk{u}(n-1),\\
\end{aligned}
\eeq

\bep
\begin{enumerate}
\item
The bracket relations \eqref{inclusion.one}--\eqref{inclusion.three} between 
the perp and parallel subspaces of $\spn$ are given as:
\begin{subequations}
\begin{align}
[\msp_\pa,\msp_\pa]= 
  & [(\a_{\pa 1},\A_{\pa 1}),(\a_{\pa 2},\A_{\pa 2})]=(\A_{\pa 2}\ov {\A}_{\pa 1}-\A_{\pa 1}\ov {\A}_{\pa 2})\in \hsp_\pa
\\
[\msp_\pa,\hsp_\pa]= 
  &[(\a_{\pa },\A_{\pa }),(\C_{\pa })]=(0,\A_{\pa }\ov {\C}_{\pa }-\C_{\pa }\A_{\pa })\in \msp_\pa
\\
[\hsp_\pa,\hsp_\pa]= 
  &[(\C_{\pa 1}),(\C_{\pa 2})]=(\C_{\pa 1}\C_{\pa 2}-\C_{\pa 2}\C_{\pa 1})\in \hsp_\pa
\\
[\msp_\pa,\msp_\pe]= 
  &[(\a_{\pa },\A_{\pa }),(\a_{\pe },\avec_{\pe })]=(2\a_\pa\a_\pe,\avec_\pe\ov {\A}_\pa -\a_\pa\ov {\avec}_\pe)\in \hsp_\pe
\\
[\msp_\pa,\hsp_\pe]= 
 &[(\a_{\pa },\A_{\pa }),(\c_{\pe },\cvec_{\pe })]=(-2\a_{\pa }\c_{\pe },\a_{\pa }\ov{\cvec}_{\pe }-\cvec_{\pe }\A_{\pa })\in \msp _\pe
\\
[\msp_\pe,\hsp_\pa]= 
 &[(\a_{\pe },\avec_{\pe }),(\C_{\pa })]=(0,\tb{a}_{\pe }\C_{\pa })\in \msp_\pe
\\
[\hsp_\pa,\hsp_\pe]= 
  &[(\C_{\pa }),(\c_{\pe },\cvec_{\pe })]=(0,-\cvec_{\pe }\C_{\pa })\in \hsp_\pe 
\end{align}
\end{subequations}
\item
The remaining bracket relations \eqref{inclusion.gen} between perp spaces are given as:
\begin{subequations}
\begin{align}
[\msp_\pe,\msp_\pe]_{\hsp_\pa}= 
 &[(\a_{\pe 1},\avec_{\pe 1}),(\a_{\pe 2},\avec_{\pe 2})]_{\hsp_\pa}
\nonumber\\
=& (\ov{\avec}^\t_{\pe 2}\ov{\wedge}\ov{\avec}_{\pe 1}
)\in\hsp_{\pa}
\label{sp.un.pe.1}
\\
[\msp_\pe,\msp_\pe]_{\hsp_\pe}= 
 & [(\a_{\pe 1},\avec_{\pe 1}),(\a_{\pe 2},\avec_{\pe 2})]_{\hsp_\pe}
\nonumber\\
=&  (2\i\Im\langle\avec_{\pe 1},\avec_{\pe 2}\rangle
,\a_{\pe 1}\ov{\avec}_{\pe 2}-\a_{\pe 2}\ov{\avec}_{\pe 1})\in\hsp_{\pe}
\label{sp.un.pe.2}
\\
[\msp_\pe,\hsp_\pe]_{\msp_\pa}= 
 &[(\a_{\pe },\avec_{\pe }),(\c_{\pe },\cvec_{\pe })]_{\msp_\pa}
\nonumber\\
=& ( -2\Re\langle\avec_{\pe},\ov{\cvec}_{\pe}\rangle
-2\a_{\pe }\c_{\pe },
\avec_{\pe}\odot\ov{\cvec}_{\pe}
)\in\msp_{\pa}
\label{sp.un.pe.3}
\\
[\msp_\pe,\hsp_\pe]_{\msp_\pe}= 
 &[(\a_{\pe },\avec_{\pe }),(\c_{\pe },\cvec_{\pe })]_{\msp_\pe}
\nonumber\\
= & (2\i\Im\langle\avec_{\pe},\ov{\cvec}_{\pe}\rangle
,\a_{\pe }\ov{\cvec}_{\pe }-\c_{\pe }\avec_{\pe })\in\msp_{\pe}
\label{sp.un.pe.4}
\\
[\hsp_\pe,\hsp_\pe]_{\hsp_\pa}= 
 &[(\c_{\pe 1},\cvec_{\pe 1}),(\c_{\pe 2},\cvec_{\pe 2})]_{\hsp_\pa}
\nonumber\\
= & (\cvec_{\pe 2}\ov{\wedge} \cvec_{\pe 1}
)\in\hsp_{\pa} 
\label{sp.un.pe.5}
\\
[\hsp_\pe,\hsp_\pe]_{\hsp_\pe}= 
 &[(\c_{\pe 1},\cvec_{\pe 1}),(\c_{\pe 2},\cvec_{\pe 2})]_{\hsp_\pe}
\nonumber\\
= & (2\i\Im\langle\cvec_{\pe 1},\cvec_{\pe 2}\rangle
,\c_{\pe 1}\cvec_{\pe 2}-\c_{\pe 2}\cvec_{\pe 1}) \in\hsp_{\pe}
\label{sp.un.pe.6}
\end{align}
\end{subequations}
\item
The Cartan-Killing form on $\msp_\pe$ is given as
\beq
\langle(\a_{\pe 1},\avec_{\pe 1}),(\a_{\pe 2},\avec_{\pe 2})\rangle
= 8(n+1)\Re(-\a_{\pe 1}\a_{\pe 2}+2\langle\avec_{\pe 1},\avec_{\pe 2}\rangle). 
\eeq
\end{enumerate}
\eep
    
The adjoint action of the Lie subalgebra 
$\hsp_\pa\subset\hsp=\un $ on  $\gsp=\spn$ generates
the linear transformation group   $H_{\pa}^\ast\subset H^\ast=\Ad(H)$, leaving invariant the element $\ee$ 
in the Cartan subspace $\mk{a}\subset\msp=\spn/\un$.
The group $H^\ast$ is given by the unitary group 
$U(n-1)\subset U(n)$ which has the matrix representation
\begin{equation}\label{nmaa}
\bpm
 1 & 0\\
 0 & C_{\pa}
\epm
\in U(n-1)\simeq H_{\pa}^\ast,
\quad
C_{\pa}\in U(n-1).
\end{equation}
This unitary group $ U(n-1)$ acts on the subspace $\msp_{\pe}$ as 
\begin{equation}\label{action.mpe} 
\Ad(C_{\pa})(\a_{\pe},\avec_{\pe})=(\a_{\pe},\avec_{\pe}C_{\pa}^t)\in\msp_\pe   .
\end{equation}
The action of $H^*_{\pa}\simeq U(n-1)$ on $\msp_{\pa}$ is given by 
\begin{equation}\label{act.h-pa-m-pa} 
\Ad(C_{\pa})(\a_{\pa},\A_{\pa})=(\a_{\pa},C_{\pa}\A_{\pa}C_{\pa}^t)\in\msp_\pa .
\end{equation}
Notice that these actions leave the scalar component unchanged. 
This observation will help us to give a geometric interpretation of 
the algebraic reduction used later to derive the SG flow. 

\begin{proposition}
The linear map 
$\ad(\ee)^2$
on $\msp_\pe,\hsp_{\pe}\simeq \i\Rnum\oplus\Cnum^{n-1}$
is given by 
$$\ad(\ee)^2(\a_\pe,\avec_{\pe})=-\chi^{-1}(4\a_\pe,\avec_{\pe})\in\msp_{\pe}, $$
$$\ad(\ee)^2(\c_\pe,\cvec_{\pe})=-\chi^{-1}(4\c_\pe,\cvec_{\pe})\in\hsp_{\pe}.$$
The irreducible subspaces of $\msp_{\pe}$ and $\hsp_\pe$
in this representation are $(\a_\pe,0)\in \i\Rnum$
and $(0,\avec_\pe)\in \Cnum^{n-1}$
on which $\ad(\ee)^2$ has respective eigenvalues
$-4/\chi$ and $-1/\chi$. 
\end{proposition}

\section{Moving parallel frames for non-stretching curve flows in Riemannian symmetric spaces }

The Riemannian structure of the space $M=G/H$ is most naturally described
\cite{KobayashiNomizu,Sharpe} in terms of
a $\msp$-valued linear coframe $e$
and a $\hsp$-valued linear connection $\conx$
whose torsion and curvature
\beq\label{tor.curv}
\mk{T}:=de+\bs{[}\conx,e\bs{]},\quad
\mk{R}:=d\conx+\ha\bs{[}\conx,\conx\bs{]}
\eeq
are $2$-forms with respective values in $\msp$ and $\hsp$,
given by the following Cartan structure equations
\beq\label{cart.stru}
\mk{T}=0,\quad \mk{R}=-\ha\bs{[}e,e\bs{]}.
\eeq
Here the brackets denote the wedge product combined with the Lie bracket. 

The underlying Riemannian metric on the space $M=G/H$ is given by 
\beq
g=-\langle e,e\rangle
\eeq
in terms of the Cartan-Killing form restricted to $\msp$. 

Now consider any smooth flow $\map(t,x)$ of a curve in $M$.
The flow is called  {\it non-stretching} if
it preserves the $G$-invariant arclength $ds=|\map_x|_gdx$,
with $|\map_x|_g^2 = g(\map_x,\map_x)$, 
in which case we can put $|\map_x|_g=1$ 
without loss of generality
(whereby $x$ is the arclength parameter). 

For flows that are transverse to the curve
(such that $\map_x$ and $\map_t$ are linearly independent),
$\map(t,x)$ will describe a smooth two-dimensional surface in $M$.
The pullback of the torsion and curvature equations \eqref{cart.stru}
to this surface yields
\begin{align}
&
D_xe_t-D_te_x+[\conx_x,e_t]-[\conx_t,e_x]=0,
\label{pull.1}\\
&
D_x\conx_t-D_t\conx_x+[\conx_x,\conx_t]=-[e_x,e_t],
\label{pull.2}
\end{align}
with
\begin{align}
&
e_x:=e\rfloor \map_x,\quad e_t:=e\rfloor \map_t, \quad
\conx_x:=\conx\rfloor \map_x,\quad \conx_t:=\conx\rfloor \map_t,
\label{pull.not.2}
\end{align}
where $D_x,D_t$ denote total derivatives with respect to $x,t$.

For any non-stretching curve flow,
these structure equations \eqref{pull.1}--\eqref{pull.not.2}
encode an explicit pair of Hamiltonian and symplectic operators
once a specific choice of frame along $\map(t,x)$ is made. 
For the case of curve flows in two-dimensional and three-dimensional 
Riemannian spaces, see Refs.~\cite{DoliwaSantini}, \cite{Sanders.Wang.Beffa}.
A proof for the general case of curve flows 
in a general Riemannian symmetric space $M=G/H$ is given in Ref.~\cite{AncoJGP}. 
See Refs.~\cite{TerngThorbergsson,MariBeffa1,MariBeffa2,MariBeffa3} 
for a related, more abstract formulation. 

As shown in Ref.~\cite{AncoJGP}, 
a natural moving parallel frame can be defined by 
the following two properties which are a direct algebraic generalization
of a moving parallel frame in Euclidean geometry\cite{Bishop}:
\newline
(i)
$e_x$ is a constant unit-norm element $\ee$ belonging to a Cartan subspace
$\mk{a}\subset \msp$, 
i.e. 
$D_xe_x=D_te_x=0$, $\<e_x,e_x\>=-1$.
\newline
(ii)
$\conx_x$ belongs to the perp space $\hsp_{\pe}$ of the Lie subalgebra
$\hsp_{\pa}\subset \hsp$ of the linear isotropy group
$H^*_{\pa}\subset H^* = \Ad(H)$ that preserves $e_x$,
i.e. $\langle \hsp_\pa,\conx_x\rangle=0$. 

A moving frame satisfying properties (i) and (ii) is called {\em $H$-parallel }
and its existence can be established by constructing \cite{AncoJGP}
a suitable gauge transformation  on an arbitrary frame
at each point $x$ along the curve. 

Through property (i), 
the set of curve flows $\map(t,x)$ in $M=G/H$ 
can be divided into algebraic equivalence classes 
defined by the orbit of the element $e_x =\ee$ 
in $\mk{a}\subset \msp$ under the action of the gauge group $H^*=\Ad(H)$.

\section{Bi-Hamiltonian structure and a hierarchy of $U(n-1)$-invariant mKdV and SG flows}

In a general symmetric space $M=G/H$, 
the Cartan structure equations \eqref{pull.1}--\eqref{pull.not.2}
yield a flow equation on the components of the Cartan connection 
along the curve,
where the flow is specified by the perp component of $\map_t$.
To write down the Hamiltonian and symplectic operators appearing 
in the flow equation, 
we use the following notation:
\begin{align}
& e_x=\ee\in\mk{a}\subset\msp_{\pa},
\\
& e_t=h_{\pa}+h_{\pe}\in\msp_{\pa}\oplus\msp_{\pe},
\\
& \conx_t=\varpi^{\pa}+\varpi^{\pe}\in\hsp_{\pa}\oplus\hsp_{\pe},
\\
& \conx_x =u\in\hsp_{\pe}. 
\end{align}
Also we write
\beq
h^{\pe}=\ad(e_x)h_{\pe}\in\hsp_{\pe}.
\eeq
Then we have the following theorem from Ref.~\cite{AncoJGP}. 
  
\begin{theorem}\label{biHam.flow.eqn}
The Cartan structure equations \eqref{pull.1}--\eqref{pull.2}
for any $H$-parallel linear coframe $e$ and linear connection $\conx$
pulled back to the two-dimensional surface $\map(t,x)$ in $M=G/H$
yield the flow equation 
\beq\label{ufloweq}
u_t = \Hop(\varpi^\pe) +h^\pe ,
\quad
\varpi^\pe = \Jop(h^\pe) 
\eeq
where
\beq\label{HJops}
\Hop = \Kop|_{\hsp_\pe} ,
\quad
\Jop= -\ad(\ee)\inv \Kop|_{\msp_\pe} \ad(\ee)\inv
\eeq
are compatible Hamiltonian and symplectic operators that 
act on $\hsp_\pe$-valued functions 
and that are invariant under $H_\pa^*$,
as defined in terms of the linear operator
\beq\label{Kop}
  \Kop := D_x +[u,\cdot\ ]_\pe -[u,D_x^{-1}[u,\cdot\ ]_\pa] .
\eeq
\end{theorem}
  
These operators arise directly from projecting 
the Cartan structure equations \eqref{pull.1}--\eqref{pull.not.2}
into the parallel and perp subspaces of $\hsp$ and $\msp$,
which yields
\begin{align}
 & D_xh_{\pa}+[u,h_{\pe}]_{\pa}=0,
\label{pull.1.th.pa}\\
 & D_xh_{\pe}+[u,h_{\pa}]+[u,h_{\pe}]_{\pe}-[\varpi^{\pe},\ee]=0, 
\label{pull.2.th.pa}
\end{align}
and 
\begin{align}
& D_x\varpi^{\pa}+[u,\varpi^{\pe}]_{\pa}=0,
\label{pull.1.pe}\\
& D_x\varpi^{\pe}-u_t+[u,\varpi^{\pa}]+[u,\varpi^{\pe}]_{\pe}+h^{\pe}=0. 
\label{pull.2.pe}
\end{align}
We will essentially use these equations \eqref{pull.1.th.pa}--\eqref{pull.2.pe}
throughout the paper.
  
As shown in Ref.~\cite{AncoJGP}, 
there are two natural flows that each give rise to a group-invariant integrable system. 
One flow is generated by the $x$-translation symmetry of the Hamiltonian and symplectic operators,
which yields a group-invariant mKdV equation. 
The other flow is defined by the kernel of symplectic operator,
which produces a group-invariant SG equation. 
These two equations are at the bottom of a hierarchy of higher-order 
integrable systems. 
  
\begin{theorem}\label{biHam.hier}
Composition of the operators $\Hop$ and $\Jop$ yields a recursion operator
$\Rop=\Hop\Jop$ that produces a hierarchy of $H^*_\pa$-invariant flows \eqref{ufloweq} 
on $u$ starting from the flow 
\beq\label{hperp.hier}
h^\pe=u_x
\eeq
which gives an integrable group-invariant mKdV equation. 
The kernel of the recursion operator $\Rop$ yields a further
$H^*_\pa$-invariant flow \eqref{ufloweq} on $u$ defined by 
\beq\label{hperp.hier.-1}
\Jop(h^{\pe})=0 .
\eeq
This flow gives an integrable group-invariant SG equation
after an algebraic reduction is made. 
\end{theorem}

The integrability structure of these two flows is shown in detail in 
Ref.~\cite{AncoJGP}.

\section{Bi-Hamiltonian flow equations in $Sp(n)/U(n)$}
\label{Sp.curveflows}

We will now derive the $U(n-1)$-invariant mKdV flow 
and $U(n-1)$-invariant SG flow
in the Hermitian symmetric space $M=Sp(n)/U(n)$. 
Employing the notation and preliminaries in \secref{g=un},
we consider a non-stretching curve flow $\map(t,x)$ that has 
a $U(n)$-parallel framing along $\map$ given as 
\begin{align}
& \ee=\frac{1}{\sqrt{\chi}}(1,0)\in\Rnum\oplus\mk{s}(n-1,\Cnum)\simeq \msp_\pa,
\quad
\chi=8(n+1)
\label{e_x.un}\\
& u=(\u,\uvec)\in\i\Rnum\oplus\Cnum^{n-1}\simeq\hsp_\pe,
\label{omega_x.un}
\end{align}
and
\begin{align}
&h_\pa=(\h_\pa,\H_{\pa})\in \Rnum\oplus\mk{s}(n-1,\Cnum) \simeq \msp_\pa,
\label{h.pa.un}\\
&h_\pe=(\h_\pe,\hvec_{\pe})\in \i\Rnum\oplus\Cnum^{n-1}\simeq \msp_\pe,
\label{h.pe.un}\\
&\varpi^\pa=(\ov{\W}^{\pa})\in \mk{u}(n-1)\simeq \h_\pa,
\label{omega_t.pa.un}\\
&\varpi^\pe=(\w^\pe,\wvec^{\pe })\in\i\Rnum\oplus\Cnum^{n-1}\simeq \h_\pe,
\label{omega_t.pe.un}
\end{align}
as well as
\beq
h^{\pe}=(\h^{\pe},\hvec^{\pe})= \ad(\ee)h_{\pe}
=\frac{1}{\sqrt{\chi}}(2\h_{\pe},-\ov{\hvec}_{\pe})
    \in\i\Rnum\oplus\Cnum^{n-1}\simeq \h_{\pe}
\label{h.pe.up.su}
\eeq
using the matrix identifications \eqref{pres.m-pa.un}--\eqref{pres.h-pa.un}, 
where 
$\u$, $\h_{\pe}$, $\w^{\pe}$, $\h^{\pe}$ $\in\i\Rnum$ 
are imaginary (complex) scalar variables,
$\h_{\pa}$ $\in\Rnum$ is a real scalar variable,
$\uvec$, $\hvec_{\pe}$, $\wvec^{\pe}$, $\hvec^{\pe}\in\Cnum^{n-1}$ $\in\Cnum^{n-1}$ 
are complex vector variables, 
$\H_{\pa}$ $\in\mk{s}(n-1,\Cnum)$ is complex symmetric matrix variable. 

We remark that, since the rank of the space $M=Sp(n)/U(n)$ is $n$, 
then for $n\geq 2$ the element \eqref{e_x.un} 
belonging to the Cartan subspace of $\msp=\mk{sp}(n)/\mk{u}(n)$
determines one particular algebraic equivalence class of 
non-stretching curve flows in which the tangent vector $\map_x$ of the curve 
is identified with the orbit of this element 
under the action of the gauge group 
$H_\pa^*=\Ad(H)$ of the $U(n)$-parallel frame. 

In terms of the variables \eqref{e_x.un}--\eqref{h.pe.up.su}, 
the Cartan structure equations \eqref{pull.1.th.pa}--\eqref{pull.2.th.pa}
and \eqref{pull.1.pe}--\eqref{pull.2.pe}
are respectively given by
\begin{align}
  &\h_{\pa x}+2\h_\pe\u+\hvec_\pe\uvec^\t+\ov{\hvec}_\pe\ov{\uvec}^\t=0,
\label{h-pa}\\
  & \H_{\pa x}-\hvec_\pe^\t\ov{\uvec}-\ov{\uvec}^\t\hvec_{\pe}=0,
\label{H-pa}\\
  &\h_{\pe x}+2\u\h_\pa+\hvec_{\pe}\uvec^\t-\ov{\hvec}_\pe\ov{\uvec}^\t-2\w^{\pe}=0,
\label{h-pe}\\
  &\hvec_{\pe x}+\uvec\H_\pa-\h_\pa\ov{\uvec}+\u_\pe\hvec_\pe-\h_\pe\ov{\uvec}+\ov{\wvec}^\pe=0,
\label{hvec-pe}
  \end{align}
and 
\begin{align}
 &\W^{\pa}_x+\ov{\wvec}^{\pe\t}\uvec-\ov{\uvec}^\t\wvec^\pe=0,
\label{W-pa}\\
 &\w^{\pe}_x-\u_t-\uvec\ov{\wvec}^{\pe\t}+\wvec^{\pe}\ov{\uvec}^\t+2\h_\pe=0,
\label{w-pe}\\
 &\wvec^{\pe}_x-\uvec_t+\uvec\W^\pa+\u\wvec^\pe-\w^{\pe}\uvec-\ov{\hvec}_\pe=0.
\label{wvec-pe}
\end{align}
Writing these equations \eqref{h-pa}--\eqref{wvec-pe} 
in the operator form \eqref{ufloweq}, 
we obtain the flow equation 
\beq\label{flow}
\bpm \u \\ \uvec\epm_t = 
\Hop\bpm\w^{\pe}\\\wvec^{\pe}\epm
+\bpm\h^{\pe}\\\hvec^{\pe}\epm,
\quad
\bpm\w^{\pe}\\\wvec^{\pe}\epm = 
\Jop\bpm\h^{\pe}\\\hvec^{\pe}\epm, 
\eeq
in terms of the Hamiltonian operator
\beq\label{Hop}
\Hop = 
\bpm 
D_x 
&\quad& 
\i 2\Im\langle\uvec,\rangle 
\\
-\uvec 
&\quad& 
D_x +\u +\uvec D_x^{-1}\uvec\ov{\wedge}
\epm
\eeq
and the symplectic operator
\beq\label{Jop}
\Jop = 
\bpm 
\tfrac{1}{4}D_x -\u D_x^{-1}\u 
&\quad& 
\i \Im\langle\uvec,\rangle +2\u D_x^{-1}\Re\langle\uvec,\rangle
\\
-\ha\uvec -\uvec D_x^{-1}\u
&\quad& 
D_x -\u +2\uvec D_x^{-1}\Re\langle\uvec,\rangle
+\ov{\uvec} D_x^{-1}\ov{\uvec}\odot
\epm .
\eeq

\subsection{mKdV flow}
The mKdV flow is produced by the $x$-translation generator
\beq\label{mkdv}
\bpm\h^{\pe}\\\hvec^{\pe}\epm
=\bpm \u_x \\ \uvec_x\epm 
=\frac{1}{\sqrt{\chi}} \bpm 2\h_{\pe}\\-\ov{\hvec}_{\pe}\epm.
\eeq
Substitution of this expression into the flow equation \eqref{flow}
yields an integrable mKdV system for the variables $(\u,\uvec)$:
\beq\label{g=sp.mkdv}
\begin{aligned}
\u_t-\chi^{-1}\u_x&=\tfrac{1}{4}\u_{xxx}-\tfrac{3}{2}\u^2\u_x
+\i3\Im\langle\uvec,\uvec_{xx}\rangle,
\\
\uvec_t-\chi^{-1}\uvec_x&=\uvec_{xxx}+\tfrac{3}{2}(|\uvec|^2-\ha\u^2)_x\uvec+3(|\uvec|^2-\ha\u^2)\uvec_x-\tfrac{3}{2}\u_x\uvec_x-\tfrac{3}{4}\u_{xx}\uvec ,
\end{aligned}
\eeq
where we have rescaled $t\rightarrow t/\chi$, for convenience. 

This system \eqref{g=sp.mkdv} has a bi-Hamiltonian structure
and exhibits invariance under the unitary group $U(n-1)$
acting on $\u$ and $\uvec$ by the transformations 
$\Ad(R)(\u,\uvec) = (\u, \uvec R^{-1})$
for all matrices $R\in H_\pa=U(n-1)$.

\subsection{SG flow}
The SG flow is defined by 
\beq\label{g=sp.un.-1flow}
0=\bpm\w^{\pe}\\\wvec^{\pe}\epm=\Jop\bpm\h^{\pe}\\\hvec^{\pe}\epm. 
\eeq
This yields the flow equation 
\beq\label{ut-flow}
\bpm\u_t\\\uvec_t\epm=\bpm\h^{\pe}\\\hvec^{\pe}\epm
=\frac{1}{\sqrt{\chi}}\bpm 2\h_{\pe}\\-\ov{\hvec}_{\pe}\epm
\eeq
with $(\h_{\pe},\hvec_{\pe})$ satisfying 
\begin{align}
  &\h_{\pe x}+2\u\h_\pa+\hvec_{\pe}\uvec^\t-\ov{\hvec}_\pe\ov{\uvec}^\t=0,
\label{-1-h-pe}\\
  &\hvec_{\pe x}+\uvec\H_\pa-\h_\pa\ov{\uvec}+\u_\pe\hvec_\pe-\h_\pe\ov{\uvec}=0,
\label{-1-tbh-pe}
\end{align}
where $\h_{\pa}$ and $\H_{\pa}$ are determined by equations \eqref{h-pa}--\eqref{H-pa}. 

Similarly to the method \cite{AncoSIGMA,AncoIMA,AncoAsadi2009,AncoAsadi2012,AncoWolf} 
used to derive SG flows in other symmetric spaces, 
we seek a local expression for $\h_{\pa}$ and $\H_{\pa}$ 
through an algebraic reduction of the form 
\beq\label{H-pa-anzats}
\H_{\pa}=\alpha\hvec_{\pe}^\t\hvec_{\pe}\in\mk{s}(n-1,\Cnum),
\eeq
with some coefficient  $\alpha(\h_{\pa},\h_{\pe})\in\Cnum$. 
Notice that, under gauge transformations \eqref{act.h-pa-m-pa}, 
it is precisely the components $\H_{\pa}$ and $\hvec_{\pe}$ that change 
while the components $\h_{\pa}$ and $\h_{\pe}$ are invariant. 
This observation provides a geometrical motivation for the form of
the reduction \eqref{H-pa-anzats}. 

To find the coefficient $\alpha$, 
we substitute expression \eqref{H-pa-anzats} into equation \eqref{H-pa} 
and use equations \eqref{-1-h-pe} and \eqref{-1-tbh-pe} to eliminate $x$ derivatives of $\h_{\pe},\hvec_{\pe}$. 
Then since the matrix $\H_{\pa}$ is symmetric, 
we expand equation \eqref{H-pa} as a linear combination of 
the symmetric matrices $\hvec^\t_{\pe}\hvec_{\pe}$ and $\ov{\uvec}^\t\hvec_{\pe}+\hvec^\t_{\pe}\ov{\uvec}$. 
Putting their coefficients to zero, we obtain 
\begin{align} 
&\alpha_x-2\alpha^2(\hvec_{\pe}\uvec^\t)-2\alpha\u=0,
\label{1-anzatz-coef}\\
&\alpha\h_{\pa}+\alpha\h_{\pe}-1=0.
\label{3-anzatz-coef}
\end{align} 
From equation \eqref{3-anzatz-coef} we obtain 
\beq\label{al-exp}
\alpha=\frac{1}{\h_{\pa}+\h_{\pe}}=\frac{\h_{\pa}-\h_{\pe}}{\h_{\pa}^2+|\h_{\pe}|^2}
\in\Cnum. 
\eeq
The remaining equation \eqref{1-anzatz-coef} is then just 
a consistency condition between equation \eqref{al-exp} for $\alpha$ 
and the equations \eqref{-1-h-pe}--\eqref{-1-tbh-pe},
which holds identically. 

Hence we have found an expression for $\H_{\pa}$ as a function of $\h_{\pa},\h_{\pe}\hvec_{\pe}$:
\beq\label{Hpa}
\H_{\pa}=\frac{1}{\h_{\pa}+\h_{\pe}}\hvec^\t_{\pe}\hvec_{\pe}=
\frac{\h_{\pa}-\h_{\pe}}{\h_{\pa}^2+|\h_{\pe}|^2}\hvec^\t_{\pe}\hvec_{\pe}.
\eeq
Next we find $\h_{\pa}$ as a function of $\h_{\pe},\hvec_{\pe}$. 
To do that we use the conservation law 
\beq\label{cons-law}
D_x(\h_{\pa}^2+|\H_{\pa}|^2-\h_{\pe}^2+2|\hvec_\pe|^2)=0
\eeq
which is admitted by the system of equations 
\eqref{-1-h-pe}--\eqref{-1-tbh-pe} and \eqref{h-pa}--\eqref{H-pa}, 
where  
\begin{align}
& 
|\H_{\pa}|^2:=\tr(\H_{\pa}\ov{\H}_{\pa})=|\alpha|^2|\hvec_{\pe}|^4=\frac{|\hvec_{\pe}|^4}{\h_{\pa}^2+|\h_{\pe}|^2}, 
\label{abs-tbHH}\\
&
|\alpha|^2=\alpha\ov{\alpha} = \frac{1}{|\h_{\pa}+\h_{\pe}|^2} = \frac{1}{\h_{\pa}^2+|\h_{\pe}|^2}.
\label{abs-al}
\end{align}
This conservation law is also given by 
$D_x( \langle h,h \rangle)=0$ 
in which $h=h_{\pa}+\h_{\pe}$ is given by equations \eqref{h.pa.un} and \eqref{h.pe.un}.
 
Substitution of expressions \eqref{abs-tbHH} and \eqref{abs-al}
into the conservation law \eqref{cons-law} gives 
\beq
|\alpha|^{-2}+|\alpha|^2|\hvec_{\pe}|^4+2|\hvec_{\pe}|^2=1 . 
\eeq 
Solving this quadratic equation for $|\alpha|^2$, we obtain 
\beq\label{al-exp2}
|\alpha|^{-2}=-(|\hvec_{\pe}|^2-\ha)\pm\ha\sqrt{1-4|\hvec_{\pe}|^2} . 
\eeq
Then from equations \eqref{al-exp} and \eqref{abs-al}
we derive the expression 
\beq\label{hpa-final}
\h_{\pa}=\pm\sqrt{|\alpha|^{-2}-|\h_{\pe}|^2}.
\eeq 

Finally, we take the $x$ derivative of the flow equation \eqref{ut-flow}
and substitute $\h_{\pa}$ and $\H_{\pa}$ given by the equations \eqref{hpa-final} and \eqref{Hpa}. 
Thus we get the following hyperbolic system for the variables $(\u,\uvec)$:
\beq
\begin{aligned}\label{SG-flow}
\u_{tx}=&-4\u A +\i4\Im\langle \uvec_t,\uvec\rangle,
\\
\uvec_{tx}=& -|\uvec|^2 \frac{A-\ha\u_t}{A^2+\tfrac{1}{4}|\u_t|^2} \uvec_t
+A\uvec-\u\uvec_t-\ha\u_t\uvec,
\end{aligned}
\eeq 
in which 
\beq
A:=\h_{\pa} = 
\pm\tfrac{1}{\sqrt{2}}\sqrt{1-2|\uvec_t|^2+\tfrac{1}{4}\u_t^2
 \pm\sqrt{1-4|\uvec_t|^2}} . 
\eeq
This system \eqref{SG-flow} is invariant under the unitary group $U(n-1)$, 
which acts on $\u$ and $\uvec$ by the transformations 
$\Ad(R)(\u,\uvec) = (\u,\uvec R^{-1})$
for all matrices $R\in H_\pa=U(n-1)$.

\section*{Conclusion}

The Hermitian symmetric space $Sp(n)/U(n)$ is one of few such spaces 
listed in the classification of symmetric spaces \cite{Helgason}. 
By adapting the moving parallel frame method developed in Ref.~\cite{AncoJGP}
to derive group-invariant bi-Hamiltonian integrable systems 
from non-stretching geometric curve flows in Riemannian symmetric spaces, 
we have obtained new integrable mKdV and SG systems for coupled 
scalar-complex vector variables. 
These systems are invariant under the unitary group $U(n-1)$ 
and have a bi-Hamiltonian structure. 

Our derivation makes use of the complex structure of the symmetric space 
$Sp(n)/U(n)$ in an essential way to formulate a natural complex matrix
representation for the spaces $\spn$ and $\un$ as well as for 
their Lie bracket relations. 
This approach resolves the problem of how to express 
the usual real matrix representation for $\spn$ and $\un$ 
in a coupled complex form that transforms properly 
under the unitary gauge group $U(n)$.
The same method can be applied to quaternionic symmetric spaces as well.

Another problem that we have addressed is how to carry out systematically 
the algebraic reduction needed to get a SG system from the underlying 
non-stretching curve flow \cite{AncoWolf,AncoSIGMA,AncoIMA,AncoAsadi2009,AncoAsadi2012}. 
We show that this reduction has a very simple geometrical formulation by 
using the projection of the parallel part of the curve flow 
into the complement of tangent direction along the curve. 
This method is motivated by our observation that the components 
given by this projection are invariant under gauge transformations
on the parallel frame along the curve. 
We are currently extending this geometrical reduction for SG curve flows 
to general symmetric spaces \cite{Anco.Asadi.Reduction}.
  
In a different direction, 
we plan to look at how to extend the work in Ref.~\cite{AncoJGP}
to derive nonlinear Schrodinger systems (complex and quaternionic) 
from non-stretching curve flows in symmetric spaces with Hermitian or quaternionic structures. 
Work is underway using examples of low-dimensional symmetric spaces
\cite{AncoAsadiAhmed}.

\section*{Acknowledgments}
EA acknowledges support from the Department of Mathematics and Statistics 
at Brock University during a research visit in which this paper was finalized.
SCA is supported by an NSERC research grant.

\end{document}